# The influence of growth interruption on the luminescence properties of Ga(As,Sb)-based type II heterostructures

Luise Rost, Jannik Lehr, Milan Maradiya, Lukas Hellweg, Florian Fillsack, Wolfgang Stolz,

and Wolfram Heimbrodt

*Department of Physics and Material Sience Center, Philipps-Universität Marburg, Germany*

**Abstract**

The influence of growth interruption on the luminescence properties of the Ga(As,Sb)/GaAs interface have been studied by continous wave and time resolved photoluminescence spectrosocopy in type II Ga(As,Sb)/GaAs/(Ga,In)As double quantum well structures. A specific highly selective etching technique in combination with atomic force microscopy (AFM) is used to analyse the morphology of the Ga(As,Sb) interface layers. The type II charge transfer recombination has been used as sensitive probe. It was found, that highest luminescence quantum efficiency can be achieved using a 10s growth interruption applying a stabilization using both precursor sources for the anion sublattice.

## 1. Introduction

The demand for high performance and efficient infrared (IR) lasers is quite high in an age where nearly every item is connected to the internet and huge amounts of data is transferred all over the world. Additionally, these laser devices get smaller, so the interface and particularly the interface morphology gets more and more important for high luminescence output [1]. So called W-lasers are a hot topic as promising candidates for infrared lasing [2–5]. These lasers are built out of semiconductors heterostructures with a type II band alignment, like Ga(As,Sb)/(Ga,As)In. The advantages are twofold, firstly there are much less Auger-losses compared to IR-lasers based on type I transitions [6,7] and secondly the structure offers more freedom in device design, because with different material combination and ternary material concentration you can tune the desired band gab To improve the wavefunction overlap the devices used to consist of two (Ga,In)As quantum layers surrounding a Ga(As,Sb) layer [8,9]. Due to the type II arrangement the laser transition is due to the carrier recombination across the interface. Therefore, interface quality and interface morphology are extremely important and actually crucial for the laser performance, since unavoidable imperfections of the interface usually act as non-radiative centers[10,11].

In a previous publication we could show that one can improve the type II photoluminescence efficiency by smoothening the interface of (Ga,In)As [12]. As extremely sensitive probe the type II recombination has been used of (Ga,In)As/GaAs/Ga(As,Sb) double quantum well (DQW) heterostructures. By applying a 120s growth interruption (GI) after growth of the (Ga,In)As layer and prior to the growth of the next layer a smoothening of the interface and enhanced luminescence efficiency was found. A detailed analysis of the other relevant layer Ga(As,Sb) and its relevance for a high quantum efficiency of the W-laser is missing so far.

It is the aim of the present paper to use the inverted DQW Ga(As,Sb)/GaAs/(Ga,In)As to analyze the influence of GI on the morphology of Ga(As,Sb) layers and the luminescence properties, particularly the recombination dynamics of the charge transfer (CT) recombination processes. We were able to reveal the strong correlation of the type II luminescence intensity and lifetime with the morphology of the Ga(As,Sb) layer and interfaces. We discuss the peculiarities of the Ga(As,Sb) using a GI in comparison to (Ga,In)As and suggest a way to get highest quantum efficiencies.

## 2. Experimental

All samples have been grown on GaAs substrate using the same growth parameters but different growth interruption times and growth conditions at the Ga(As,Sb) surface. The samples were grown in a commercial AIXTRON AIX200 GFR (gas foil rotation) metal organic vapor phase epitaxy (MOVPE) reactor system. The group-III precursors triethylgallium (TEGa) and trimethylindium (TMIn) were used in combination with triethylantimony (TESb) and the arsenic precursor tertiarybutylarsine (TBAs) for the epitaxial growth. The reactor pressure was set to 50 mbar with $H_2$ as carrier gas. A TBAs-stabilized bake-out procedure was applied in order to remove the native oxide layer from the semi insulation GaAs (001)(±0,15°) substrate surface. The growth of the samples was carried out at 550 °C.

A schematic picture of the DQW structures is given in figure 1. An interface modification was achieved by a GI of either 10s or 120s after the Ga(As,Sb) layer was grown. The positions of the GI is depicted in yellow in figure 1. We prepared two sets of samples. For the sample set (1) the TBAs partial pressure used to grow the layer was held constant over the GI time, but the triethylantimon (TESb) and the triethylgallium (TEGa) sources were closed. For the second set TBAs and TESb sources were kept open. For comparison a sample was grown without any GI. The active region of the samples consists of a 5× DQW heterostructures.

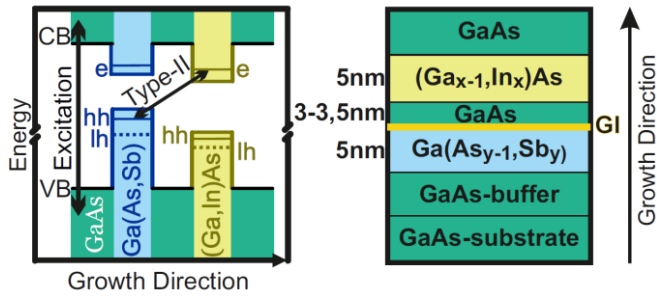

*Figure 1* Sample structure of type II heterostructures (right) and schematic depiction of the electronic band structure (left). The arrows indicate the excitation and type-II transition, respectively. Heavy hole and electron states in the quantum wells are depicted schematically as thin solid lines. The light hole states (dotted lines) are shifted due to strain and confinement.

Each DQW is composed of a 5 nm thick $Ga(As_{1-x},Sb_x)$ layer which is followed by a GaAs interlayer of 3.5 nm thickness. The active region is completed by a second 5.0 nm thick quantum well consisting of $(Ga_y, In_{1-y})As$ which is followed by a 50 nm thick GaAs barrier. The structures have been characterized by high resolution X-Ray diffraction (HRXRD). [2,13]( cf. page S-2). The determined In and Sb concentrations and layer thicknesses are summarized in table 1, respectively.

We used the continuous wave photoluminescence (cw-PL) and time resolved photoluminescence (TRPL) to analyze the luminescence properties. The cw-PL spectra were measured using a liquid nitrogen cooled Ge-detector and a 0.25m grating spectrometer. A frequency-doubled Nd:YAG laser at 532 nm in cw-mode provided the light for the excitation of the samples. The time-resolved PL measurements were performed using a frequency-doubled Nd:YAG at 532nm in pulsed mode at a repetition rate of 10 Hz and a full width at half maximum of the laser pulse about 3 to 4 ns. The PL was detected using a 0.25 m grating spectrometer and a thermoelectrically cooled (In,Ga)(As,P)/InP photomultiplier, with a time resolution of 2ns

*Table 1* In and Sb concentrations and layer thicknesses d of the investigated heterostructures determined by analysing the HR-XRD spectra. The error-bars for the given compositions are ±0,5% for each material and ±0,2nm for each of the respective layer thicknesses d.

| Growth interruption time | GaAsSb | | GaAs | GaInAs | |
|---|---|---|---|---|---|
| | Sb [%] | d [nm] | d [nm] | In [%] | d [nm] |
| **0s** | 23,5 | 4,8 | 3 | 19 | 5 |
| **10s** *TBAs* | 23 | 4,5 | 3,5 | 19 | 5 |
| **120s** *TBAs* | 22,5 | 4,4 | 3,6 | 19 | 5 |
| **10s** *TBAs+TESb* | 23,5 | 4,8 | 3,3 | 20 | 5 |
| **120s** *TBAs+TESb* | 23,5 | 4,5 | 3,5 | 20 | 5 |

To analyse the interior interface morphology, we used a highly selective etching procedure[14,15]. Ga(As,Sb) layers have been grown at 525 °C and capped with a chemically sufficiently different layer of AlAs. Afterwards, one can wet chemically remove the capping layer. Hydrofluoric acid removes the AlAs layer without affecting the GaAsSb layer. Hence, an analysis of the real interior interface as it was grown is possible by means of subsequent atomic force microscopy (AFM). It is expected, that morphological changes are slightly different for different capping layers (see e.g. [16]), but trends and tendencies can be revealed clearly. Therefore, we were able to compare the changes in interface morphology with and without growth interruption (GI) procedures.

### 3. Results and Discussion

In figure 1 the band structure for the Ga(As,Sb)/GaAs/(Ga,In)As DQWs is depicted schematically. The arrows indicate the optical excitation and the electron-hole recombination, e.g. the CT transitions, respectively. The holes are confined in the Ga(As,Sb) well and the electrons are in the (Ga,In)As well [2,17]. The type-II transition can be tuned to lower transition energies in these DQWs either by the strong valence band shift of the Ga(As,Sb) well to higher energies with higher Sb concentrations or by a higher In content and respective conduction band shift to lower energies in the (Ga,In)As well[18,19].

We now want to study the influence of modified interfaces on the CT recombination. We used GI at the interface between Ga(As,Sb) and the inner GaAs barrier during the MOVPE growth of our structures. In the first run the GI was carried out as described before, under constant TBAs partial pressure. The GI time was 10s and 120s, respectively. In figure 2 the AFM images of the Ga(As,Sb) interior interface after GI of 10s and 120s (only TBAs stabilized) are compared with an as-grown interface without GI.

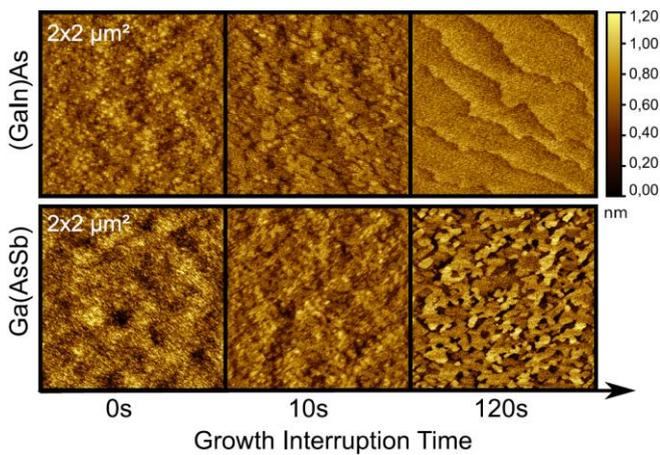

*Figure 2 Atomic Force Microscopy Images after selective etching to the interior interface of the respective layer with only TBAs present during GI time.*

For comparison there are shown also the AFM images of (Ga,In)As surfaces in figure 2, which have been achieved using the same GI and etching procedure. (Ga,In)As shows the same behaviour known for GaAs [20]. The growth interruption actually leads to the formation of big, smooth terraces, which starts already with 10s GI and is fully developed with GI of 120s.

For the Ga(As,Sb) the AFM images look completely different, instead of smooth terraces small islands with steep edges and deep holes are formed (c.f. fig. S1+S2).

The HR-XRD results show (i) a thinner Ga(As,Sb) QW with longer GI duration compared to the reference sample, (ii) additionally a slightly reduced Sb concentration and (iii) GaAs interlayers, whose thicknesses increases with longer GI. It should be mentioned at this point, that the growth rate and growth time, like every other growth parameter was always kept the same. We have to conclude from these facts, that the Sb atoms desorb from the growth surface during GI but also from the already grown layer itself.

As a result we get an effectively thinner Ga(As,Sb) and an effectively thicker GaAs layer. We come back to this point later. The deep holes in the interface seen in figure 2 for GI of 120s are obviously a result of the missing Sb atoms. The observed surface in the AFM picture should therefore consist of GaAs terraces with holes reaching into the Ga(As,Sb) QW.

In the set of samples with TBAs and TESb stabilisation during the GI, it can be seen from the HR-XRD results in table 1, that these losses are avoided. The HR-XRD results of these samples resemble much better the parameters of the reference sample.

In the following we want to correlate the observed structural changes with the luminescence properties of the complete heterostructures. In figure 3 the cw-PL spectra taken at T=300K are depicted for the reference sample (0s GI in black) and samples with 10s and 120s GI with TBAs and without TESb partial pressure during GI, respectively. The intensity of the PL varies slightly across the sample, mainly due to concentration fluctuations typical for ternary materials. That's why three laterally different positions on the sample were measured and the mean value is shown in figure 3. The shaded areas indicate the respective intensity variations (cf. fig. S3). The energetically lower energy band is due to the type II recombination. At room temperature also the type I transition of the Ga(As,Sb) well at about 1.1eV can be seen. The small conduction band (CB) offset between the (Ga,In)As and Ga(As,Sb) well allows for a thermal population of both CB wells. In so far, the spatially indirect and the direct recombination take place. It should be mentioned at this point, that the type I transition in the (Ga,In)As well is not observable at room temperature due to an effective hole tunnelling into the Ga(As,Sb) well (see [21]).

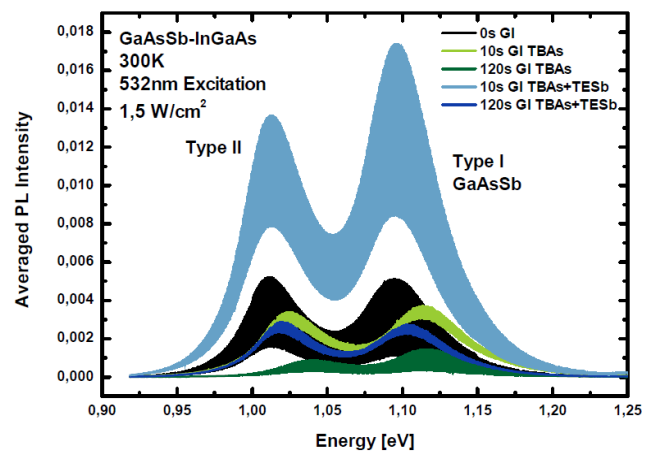

*Figure 3 Mean photoluminescence of the (Ga,In)As/GaAs/Ga(As,Sb) heterostructures taken at three different spatial positions, respectively.*

As can be seen in figure 3 the PL intensity changes substantially from sample to sample but also the peak maximum energy is shifted to higher energies in case of GI with TBAs and without TESb partial pressure. A relatively strong shift to higher energies can be seen for the direct Ga(As,Sb) PL band for the samples with 10s GI (light green) and 120s GI (dark green) with only TBAs partial pressure compared to the reference sample (black). There are two obvious reasons: (i) the confinement shift due to the reduced well thickness (see table 1) and (ii) the reduced Sb concentration as mentioned before which enhances the bandgap energy. The underlying shift of the valence band explains the respective blueshift of the type II transition.

The PL intensity of the direct and type II transition for the sample with 120s GI with only TBAs is tremendously reduced. This is explained by the extremely rough interface as source of a high number of non-radiative centers. There is no blueshift and a very high PL intensity for the sample with 10s GI TBAs and the TESb precursor being present additionally. This is a clear hint that the Sb loss can be avoided this way and the number of non-radiative centers is substantially reduced in comparison to the reference sample.

In the following we will have a look on the influence of the modified interfaces on the electron-hole recombination time through the GaAs barrier. We performed time resolved PL analysis of the type-II transition. In figure 4 the decay curves are depicted for the type-II PL of all the DQWs at room temperature. The transients were measured at the maximum of the type-II PL band and are normalized to unity.

The decay curves exhibit almost a mono-exponential behaviour. Little deviations from the mono-exponential decay at late times are caused by feeding processes of the type-II excitons which we discussed earlier [21]. The initial decay is the effective type-II lifetime. All lifetimes are in the ns range which is typical for type-II transitions[17].

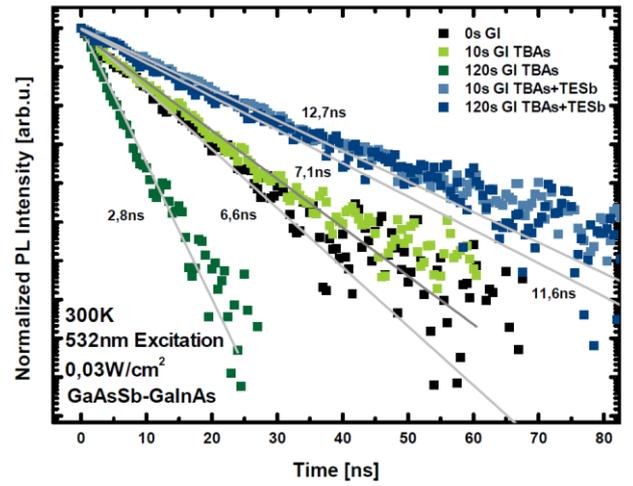

*Figure 4* Room-temperature decay curves of the type-II PL with and without GI. The curves were normalized to unity and were taken at the maximum of the type II peak. The solid curves are fitted transients using a single-exponential decay.

The reference sample has a decay time of 6.6 ns. The sample with 10s GI with TBAs partial pressure (light green) exhibits nearly the same decay time of 7ns but the sample with 120s GI (dark green) with TBAs shows a fast decay time of 2,8ns which is actually at the limit of our temporal resolution. The result is in total agreement with the cw-PL data discussed before. The strong enhancement of non-radiative centers accelerates the decay.

This is underlined also by the PL-decay of the samples with a GI under TBAs and TESb. Both samples exhibit an increasing recombination time. The decay time of the sample with 10s GI TBAs+TESb (light blue) is slightly longer with 12,7 ns than for that with 120s GI TBAs+TESb (dark blue) with 11,6 ns. This is in agreement with the higher cw-PL intensity of the sample with 10s GI.

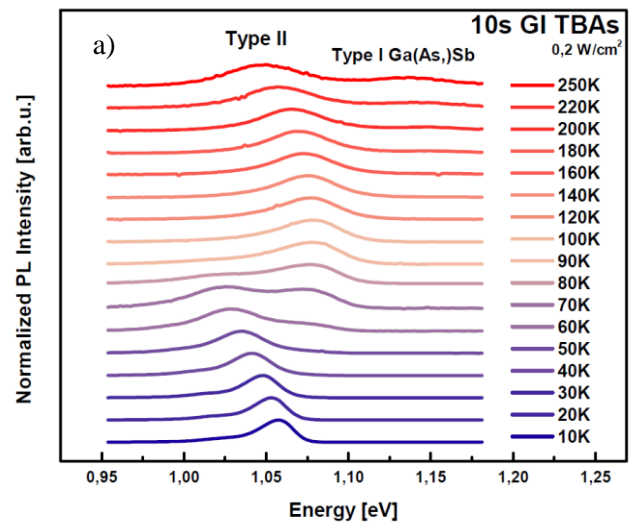

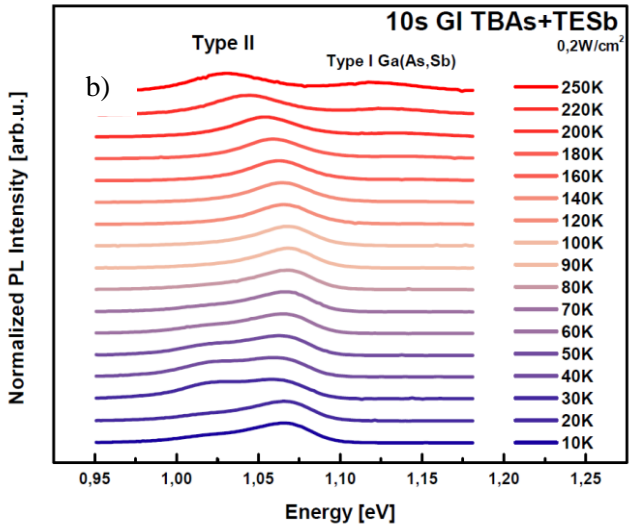

*Figure 5 cw-PL spectra in the temperature range T=10K to 250K for the samples with 10s GI (a) with TBAs and (b) with TBAs+ TESb for Sb stabilization. The spectral maxima are normalized to unity and shifted relatively to each other.*

It is known, that microscopic disorder can be characterized by temperature dependent PL measurements[22–24]. Disorder causes due to electronic potential fluctuations a huge amount of strongly localized states within the band gap. This can be observed by a non-monotonous emission energy shift with rising temperature, the so called s-shape. This can be described by a model in which excitons were able to travel through the spatially localized states by hopping transport[25,26]. To correlate the interface morphology and charge transfer properties we measured the cw-PL in the temperature range from 10K up to 250K. In figure 5 a) and b) the cw-PL spectra are shown as examples at various temperatures for the samples with 10s GI with TBAs (fig5a) and with TBAs and TESb (fig5b). The temperature was varied from 10K to 250K and the PL maximum was normalized and shifted for better visualisation. For the measurements a low excitation power (0,2W/cm^2) was chosen to realize low exciton densities, which is necessary to make sure that the disorder induced S-shape can be observed. In Figure 5a) at low temperatures the typical PL band caused by localized excitons can be seen, which vanishes for temperatures above 80K. A second peak rises at about 50K and can be measured up to room temperature. At an intermediated temperature regime around 70K both PL bands are observable in parallel. It should be mentioned at this point, that for higher excitation power both bands merge into one with higher half width. It is important to note, that both PL bands are caused by spatially indirect type II recombination. All spatially direct excitons have much higher transition energies [21]. In the exciton picture the two PL bands are due to the localized and free excitons, respectively.

In Figure 5b the very interesting result can be seen for the sample with TBAs and TESb present during 10s GI. The free exciton can be measured down to 10K and the PL band of the localized states is very weak and only noticeable between 30 and 50K. A strongly reduced number of localized states yield a fast filling and saturation and the free excitons are observable at all temperatures.

In Figure 6a and 6b the peak position energy are plotted versus temperature for all the samples with TBAs and TBAs+TESb during growth interruption in comparison to the reference sample. It can be seen clearly, that both peaks behave completely different with temperature. The lower energy peaks show a minimum which is clearly the well-known S-shape behaviour due to the temperature dependent hopping mobility of localized excitons in disorder induced states. The high energy peak exhibits the typical Varshni-behaviour [27]. The recombination energy of the free excitons decreases with increasing temperature.

The critical temperature which is determined by the minimum is a measure of the disorder in our samples. For the reference sample we detect a $T_c$=30K. The GI with TBAs only strongly increases $T_c$ to 65K in case of 10s GI and $T_c$=80K with 120s GI. The increased microscopic disorder is in accordance with the AFM data. It is interesting to note at this point, that the underlying disorder potential is in the Ga(As,Sb) well, i.e. the results resembles the fluctuations of the hole potential.

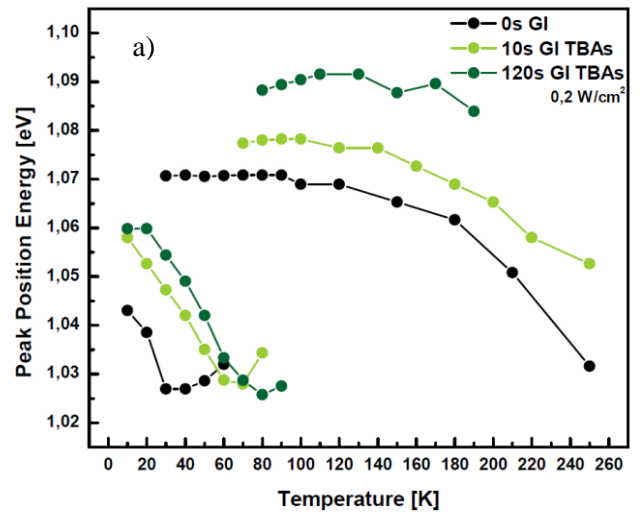

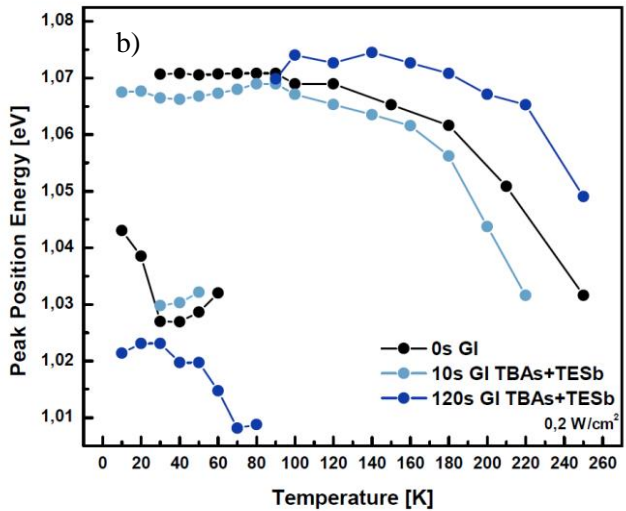

*Figure 6 Peak Position Energy over temperature. The connections between the different temperatures are only a guide to the eye. a) Depicted in green are the samples with 10s and 120s GI TBAs and b) the samples with additionally TESb during 10s and 120s GI.*

In case of TBAs and TESb stabilization during GI the results are different (see Fig.6b). For the sample with 120s GI a minimum is hardly visible, but $T_c$ is about 80K. In case of 10s GI a minimum cannot be seen, but from the energy shift of the localized states we have to conclude, that it is below 30K, i.e. less disorder in perfect agreement with the high luminescence efficiency.

## 4. Conclusion

In summary we studied the luminescence properties of Ga(As,Sb)/GaAs/(Ga,In)As type II double quantum wells in dependence on different growth interruption (GI) procedures.

We were able to reveal a correlation of the morphology of Ga(As,Sb) interfaces and charge transfer PL intensity and lifetime.
GI with only TBAs stabilization results in a substantial Sb loss, as was observed by combining HR-XRD and AFM analysis, yielding a rough Ga(As,Sb) interface, deep holes and even an effectively reduced Ga(As,Sb) layer thickness. This results in a substantially reduced type II PL intensity and recombination time at room temperature with increasing GI time. The morphology changes due to the Sb loss cause an increasing number of non-radiative centers.
It was found, that this deterioration of the interface could be overcome with Sb precursor TESb present during the GI additionally. This suppresses the sublimation loss of Sb atoms and results in smoother interfaces with strongly reduced number of non-radiative centers. The cw-PL intensity is substantially enhanced using 10s GI. Even the type II recombination time is enhanced due to the supressed loss channel. The achieved high quality of the Ga(As,Sb)/GaAs interface could be convincingly shown by the temperature dependent PL measurements. The Type II recombination of the non-localized carriers and excitons could be observed down to lowest temperatures. The localized carriers could be hardly seen, since the number of hole trap states is obviously substantially reduced. It is important to note, however, that there is an optimal GI-time. A too long GI leads again to a degradation of the interface.

**Acknowledgements**
We are grateful for financial support by the German Science Foundation (DFG) in the framework of the SFB 1083.